\documentclass[11pt]{article}
\usepackage{color}
\usepackage{epsfig}
\usepackage{subfig}
\usepackage{mathrsfs}
\usepackage{booktabs}
\usepackage{graphicx}
\usepackage{epstopdf}
\usepackage{multirow}
\usepackage{amsmath,amssymb}
\usepackage{amsthm}
\usepackage{fancyhdr}
\usepackage{tabularx}
\usepackage{xfrac}
\usepackage{authblk}
\usepackage{algorithm}  
\usepackage{algorithmic}  

\newtheorem{remark}{Remark}[section]

\begin{document}

\title{Identifying Critical Risks of Cascading Failures in Power Systems}

\author[1,2,3]{Hehong Zhang}
\author[1,2,4]{Chao Zhai}
\author[1,2,4]{Gaoxi Xiao\thanks{EGXXiao@ntu.edu.sg}}
\author[1,2]{Tso-Chien Pan}
\affil[1]{{\small Future Resilient Systems, Singapore-ETH Centre, CREATE Tower, Singapore}}
\affil[2]{{\small Institute of Catastrophe Risk Management, Nanyang Technological University, Singapore }}
\affil[3]{{\small Interdisciplinary Graduate School, Nanyang Technological University, Singapore}}
\affil[4]{{\small School of Electrical and Electric Engineering, Nanyang Technological University, Singapore}}

%\author{Hehong Zhang, Chao Zhai, Gaoxi Xiao and Tso-Chien Pan
%\thanks{Hehong Zhang is with Interdisciplinary Graduate School, Nanyang Technological University, Singapore. Chao Zhai and Gaoxi Xiao are with School of Electrical and Electric Engineering, Nanyang Technological University, Singapore. Tso-Chien Pan is with Institute of Catastrophe Risk Management, School of Civil and Environmental Engineering, Nanyang Technological University, Singapore. All authors are also with Future Resilient Systems, Singapore-ETH Centre, 1 CREATE WAY, CREATE Tower, 138602 Singapore. Corresponding author: Gaoxi Xiao, Email: EGXXiao@ntu.edu.sg.}}

\maketitle

\begin{abstract}
Potential critical risks of cascading failures in power systems can be identified by exposing those critical electrical elements on which certain initial disturbances may cause maximum disruption to power transmission networks. In this work, we investigate cascading failures in power systems described by the direct current (DC) power flow equations, while initial disturbances take the form of altering admittance of elements. The disruption is quantified with the remaining transmission power at the end of cascading process. In particular, identifying the critical elements and the corresponding initial disturbances causing the worst-case cascading blackout is formulated as a dynamic optimization problem (DOP) in the framework of optimal control theory, where the entire propagation process of cascading failures is put under consideration. An Identifying Critical Risk Algorithm (ICRA) based on the maximum principle is proposed to solve the DOP. Simulation results on the IEEE 9-Bus and the IEEE 14-Bus test systems are presented to demonstrate the effectiveness of the algorithm. 
\end{abstract}

\emph{\textbf{Index Terms}}---Cascading failures, critical elements, initial disturbances, dynamic optimization, maximum principle.

\section{Introduction}
Almost all human systems and activities strongly depend on critical energy infrastructures (e.g., electric power systems). Large-scale power blackouts in the past decades, such as the North America blackout on August 14, 2003 \cite{1}, the Europe interconnected grid blackout on November 12, 2006 \cite{2} and Brazil blackout on November 10, 2009 \cite{3}, suggest that power blackouts are not uncommon in spite of technological progress and great investments in power systems \cite{4}. Although such large blackouts are rare events, they have the potential to result in in-operability, huge economic losses, or even state panics. Cascading failures in bulk power systems are an essential cause of blackouts \cite{5}. A cascading blackout usually starts with one or more triggering initial disturbances that lead to dramatic redistributions of power flows and a variety of drastic phenomena throughout the power network \cite{6}. Therefore, identifying critical risk of cascading failures in power systems is of great interest to researchers and power system planners. Certain disturbances on some elements may lead to the worst power losses or severest isolations of power systems, making these elements the critical elements for cascading failures in power systems \cite{7}. To identify the critical elements and the corresponding initial disturbances applied on them that cause the worst-case cascading blackout in power systems, a novel approach within the framework of optimal control theory is proposed in this paper.

A variety of approaches have been proposed to identify critical electrical elements and initial attacks, or to assess the criticality or vulnerability for power systems. In \cite{8}, identifying critical system components (e.g., transmission lines, generators, transformers) is formulated into a bi-level optimization model; and a heuristic algorithm is developed to solve the problem and obtain a local optimal solution. In \cite{9}, the problem is recast into a standard mixed-integer linear programming problem, which can be solved by using various solvers. The resulted mixed-integer bi-level programming formulation in \cite{8}\cite{9} is relaxed into an equivalent single-level mixed-integer linear programming problem by replacing the inner optimization problem with the Karush-Kuhn-Tucker optimality conditions \cite{10}. As an extension of \cite{8}, a new approach based on ``Global Benders Decomposition" is proposed to solve the large-scale power system interdiction problem when transmission lines are under attacks; and the algorithm can guarantee the convergence of the bi-level optimization solution \cite{11}. In \cite{12}, identifying the criticality and vulnerability of the electric grid is formulated as a non-linear bi-level programming problem and the genetic algorithm is appiled to reach near optimal solutions with moderate computing time. In \cite{13}, finding a strategic defense to minimize the damages of an attack is formulated as a multi-level mixed-integer programming problem. A Tabu Search with an embedded greedy algorithm is implemented to find the optimum defense strategy. In \cite{14}, an improved interdiction model that combines the evaluation of  both short-term (seconds to minutes) and medium-term (minutes to days) impacts of possible electric grid attacks to identify the worst one is proposed; an integer programming heuristic is then applied to solve the problem. Power grid performance indices including overall voltage deviation and the minimal load shedding are quantified in \cite{15} based on the alterinating current (AC) power flow model, where finding the most disruptive attack is formulated as either a non-linear programming or a non-linear bi-level optimization problem, both of which can be solved by common algorithms. In \cite{16}, both static and dynamic deterministic indices are included in the process of ranking critical nodes; a new ranking algorithm is proposed and evaluated by extensive Monte Carlo simulations. 

In most of the existing work on identifying the critical elements and the initial disturbances causing the worst-case cascading blackout, the problem is usually formulated as a static optimization problem which neglects the entire propagation process of the cascading failures. Though such a problem is relatively easier to be solved, the results however may be misleading as they may not properly reflect the system dynamics and evolution in the real life. 

The main contributions in this paper are twofold. Firstly, we formulate the problem of identifying the critical elements and the corresponding initial disturbances causing the worst-case cascading blackout as a dynamic optimization problem (DOP) in the framework of optimal control theory, which enables us to investgate the entire propagation process of cascading failures. Secondly, the identifying critical risk algorithm (ICRA) based on the maximum principle of optimal control theory \cite{17}\cite{18}\cite{19} is proposed to solve the DOP, which guarantees fulfilling the necessary condition for the optimal solutions. 

The remainder of this paper is organized as follows. Section 2 formulates the DOP based on the DC power flow equations and cascading failure model. In Section 3, the solution based on the maximum principle is introduced in detail. Section 4 presents results from calculations based on the IEEE standard data and verifies the correctness of the results. Finally, we conclude this work and present some future work in Section 5.

\section{Problem Formulation}\label{sec:problem}
In this section, identifying the critical elements and the corresponding initial disturbances is formulated as a dynamic optimization problem (DOP). The DC power flow model, relay-based overloading branch tripping model and cascading failure model are discussed in Section 2.1, and the DOP formulation is presented in Section 2.2.

\subsection{Notations and Models}

\subsubsection*{A. Notations}
 We summarize the power system notations used in later sections as follows:

\begin{itemize} \item Number of buses: $N_{b}$\end{itemize} 
\begin{itemize} \item Number of electrical elements: $N$\end{itemize}
\begin{itemize} \item Active power at bus $i: P_i$\end{itemize}
\begin{itemize} \item Active power from bus $i$ to bus $j: P_{ij}$\end{itemize}
\begin{itemize} \item Voltage phase at bus $i : \theta_i$\end{itemize}
\begin{itemize} \item Voltage phase difference between bus $i$ and bus $j$: $\theta_{ij}$\end{itemize}
\begin{itemize} \item Admittance at element $i: y_{pi}$\end{itemize}

The admittance of an element includes admittance of transformer (if any) and transmission branch. The admittance information of a power system can be described by the element admittance vector $Y_P=[y_{p1}\quad y_{p2} \quad \cdots \quad y_{pN}]^T$. An initial disturbance is specified by means of altering admittance at the corresponding element of $Y_P$. The nodal admittance matrix $Y$ can be determined by $Y=A^TY_PA$, where $A$ is the element-node incidence matrix \cite{20}\cite{21}. In the propagation process of cascading failures, the time-varying element admittance vector $Y_P$ and the time-invariant element-node incidence matrix $A$ are applied to determine the nodal admittance matrix $Y$ for the convenience of analysis of the approach in later sections.

\subsubsection*{B. DC Power Flow Model }
In a power system, power flow equations are used to estimate the flow values for each branch. The DC power flow model is deployed since we only study on high-voltage transmission networks in this paper: adopting the DC model helps avoid some difficulties in numerical calculations without sacrificing the validity of the results \cite{22}.

In the AC power flow model, the active power flow $P_{ij}$ is determined as:
\begin{equation}
P_{ij}=\frac{|U_i||U_j|}{z_{ij}}{\rm sin}\theta_{ij}
\end{equation}
where $|U_i|$ is the voltage amplitude at bus $i$, $z_{ij}$ is the element impedance. Under the following assumptions that i) resistance of transmission element is ignored so that element impedance approximately equals element reactance; ii) voltage phase differences are small enough and iii) there is a flat voltage profile \cite{23}, the above non-linear equation can be linearised into the DC power flow equation.
\begin{equation}
P_{ij}=\frac{\theta_{ij}}{x_{ij}}=y_{ij}\theta_{ij}
\end{equation}
Further, the power flow equation can be modelled into matrix format as follows:
\begin{equation}
P=A^TY_PA\theta
\end{equation}
where $P$ is the vector of active power injections, vector $\theta$ contains the voltage angles at each bus, and $A^TY_PA$ is the nodal admittance matrix $Y$. Due to ignorance of power loss in the DC power flow equations, all the active power injections are known in advance. Once given the nodal admittance matrix, the voltage angles at each bus can be determined by
\begin{equation}
\theta=(A^TY_PA)^{-1}P
\end{equation}
After obtaining the voltage angle value at each bus, the power flow through each element can be computed by Eq.\,(2).

\subsubsection*{C. Relay-Based Overloading Branch Tripping Model}
In a power system, transmission branches are protected by circuit breakers, and branch tripping is one of the most common factors responsible for cascading failures. A circuit breaker trips a transmission branch when the demand load of the branch exceeds a certain threshold level, in order to prevent that transmission branch from being permanently damaged due to overloading \cite{24}.

For simplicity, in this paper we assume a deterministic model of transmission branch tripping mechanism. Specifically, a circuit breaker for branch $l_i$ trips at the moment when the demand load on the branch $l_i$ exceeds its maximum capacity (threshold value). The maximum capacity of a branch is defined as the maximum power flow that can be afforded by the branch. This maximum power flow value is decided by thermal, stability and/or voltage drop constraints. In real-life infrastructures, this value may be constrained by cost as well. The relay-based overloading branch tripping model is presented as follows, where the threshold value of a branch is related to its initial load:
\begin{equation}
C_{tr1i}=\alpha_{i}L_{i}(0) \quad\quad i=1,2,...,N
\end{equation}
where $L_{i}(0)$ is the initial demand load, and $\alpha_{i}$ is the tolerance parameter of line $l_i$.

The mechanism of the relay protection above may be resembled by a step function: when the real load of a branch is less than or equal to the threshold value, the circuit breaker of the branch is in the status of $on$; otherwise, it is in the status $off$. To facilitate derivative calculations, a smooth function $g$ is introduced to resemble the step function, ensuing the differentiability of the function at switching points: 
\begin{eqnarray}g_{i}(p_{ij},C_{tr1i})=
\begin{cases}
1, &|p_{ij}|\leq \sqrt{C_{tr1i}^{2}-\frac{\pi}{2a}}\cr \frac{1-{\rm sin}a(|p_{ij}^{2}|-C_{tr1i}^{2})}{2}, &\sqrt{C_{tr1i}^{2}-\frac{\pi}{2a}}\leq |p_{ij}| \leq \sqrt{C_{tr1i}^{2}+\frac{\pi}{2a}} \cr 0, &|p_{ij}|\geq \sqrt{C_{tr1i}^{2}+\frac{\pi}{2a}} \end{cases}
\end{eqnarray}
where $C_{tr1i}(i=1,2,...,N)$ is the threshold value of a branch, $a$ is a parameter to regulate the slope of the function. With the smooth function $g_{i}(p_{ij},C_{tr1i})$, the diagonal relay tripping matrix $G(p_{ij},C_{tr1i})$ can be defined as follows:
\[
\text G(p_{ij},C_{tr1i}) =
\begin{bmatrix}
g_{1}(p_{ij},C_{tr11})  \\
 &g_{2}(p_{ij},C_{tr12})  \\

 & &  & \ddots & \\
 &  &  &  & g_{N}(p_{ij},C_{tr1N})
\end{bmatrix}
\]

\subsubsection*{D. Cascading Failure Model}

In this subsection, the cascading failure model reflecting the entire propagation process of cascading failures is presented. A cascading failure is a sequence of events in which an initial disturbance, or a set of disturbances, triggers a sequence of one or more dependent element outages. The initial disturbances include a wide variety of exogenous disturbances such as high winds, lightning, natural disasters, contact between conductors and vegetation, or human errors \cite{25}. For simplicity, we assume that the initial disturbances take the form of altering admittance along transmission branches.

From Eq.\,(6) and the diagonal relay tripping matrix $G(p_{ij},C_{tr1i})$, the cascading failure model in matrix format can be built as follows:
\begin{equation}
Y_P^{k+1}=G[P_{ij}^{k}(Y_P^{k}),\ C_{tr1}]Y_P^{k}+Diag[-u(k)]F(u_{k})\quad k=0,1,2,...
\end{equation}
where $k$ is the iteration step of cascading failures, $u(k)$ is the input vector of external disturbances. When $k=0$, the input vector $u(0)$ denotes the initial disturbances. The vector $F(u_{k})$ is defined as follows: 
\[
\text F(u_{ki},C_{tr2i}) =
\begin{bmatrix}
f_{1}(u_{k1},C_{tr21})  \\
 f_{2}(u_{k2},C_{tr22}) \\

  \vdots & \\
f_{N}(u_{kN},C_{tr2N})
\end{bmatrix}
\]
Similar to that in Eq.\,(6), for facilitating derivative calculations, a smooth function $f_i(u_{ki},C_{tr2i})$ is applied for every element of the vector $F(u_{k})$. The smooth function $f_i(u_{ki},C_{tr2i})$ is defined as follows:

\begin{eqnarray}f_i(u_{ki},C_{tr2i})=
\begin{cases}
0, &|u_{ki}|\leq \sqrt{C_{tr2i}^{2}-\frac{\pi}{2b}}\cr \frac{1+{\rm sin}b(|u_{ki}^{2}|-C_{tr2i}^{2})}{2}, &\sqrt{C_{tr2i}^{2}-\frac{\pi}{2b}}\leq |u_{ki}| \leq \sqrt{C_{tr2i}^{2}+\frac{\pi}{2b}} \cr 1, &|u_{ki}|\geq \sqrt{C_{tr2i}^{2}+\frac{\pi}{2b}} \end{cases}
\end{eqnarray}
where $C_{tr2}$ is the threshold value vector and $b$ is a parameter that regulates the slope of the function $f$. The returning value of the function $f_{i}(u_{ki},C_{tr2i})$ is determined by comparing the threshold value $C_{tr2i}$ with the corresponding external disturbance, where the critical element is determined when the function $f$ returns the value one.

\subsection{Dynamic Optimization Problem Formulation} 

Based on the models presented above, the DOP formulation in the framework of optimal control theory can be defined as follows:

\textbf{Formulation of DOP:} Given a power system, determining a control input vector ${u_{k}\in \Omega}$, such that the remaining transmission power at the end of cascading process is minimized. Assume that the system is described by the DC power flow equations in Eq.\,(2) and its cascading failure model by Eq.\,(7). We have
\begin{eqnarray}
&\min\limits_{u_{k}\in \Omega}\quad J\\
&J=||P^{N}||_{F}^{2}+\epsilon\sum\limits_{k=0}^{N_c-1} [\frac{1}{\max\left\{0,1-k\right\}}\times\frac{1}{\max^2\left\{0,N_n-||F(u_k)||^{2}\right\}}]\\
& \begin{array}{r@{\quad}r@{}l@{\quad}l}
\text{s.t.} & \left\{
\begin{aligned}
&Y_P^{k+1}=G[P_{ij}^{k}(Y_P^{k}),\ C_{tr1}]Y_P^{k}+Diag[-u(k)]F(u_{k}) \\
&P_{ij}=y_{ij}\theta_{ij} \\
\end{aligned}
\right.
\end{array} 
\end{eqnarray}
where the Frobenius norm of power transmission matrix $||P^{N}||_{F}^{2}$ equals ${{\sum\limits_{i=1}^N\sum\limits_{j=1}^N}(P^{N}_{ij})^2}$, $N_c$ is the total iteration steps of cascading failures, $N_n$ is a parameter that denotes the number of critical elements, $\|\cdot\|$ denotes 2 norm of a vector and $\epsilon$ is the weight of the cost function. In Eq.\,(10), the terminal constraint $||P^{N}||_{F}^{2}$ dominates the cost function by setting the weight $\epsilon$ to be small enough.

\begin{remark}
As above mentioned, the critical elements are those elements which, when being attacked, will trigger the worst-case cascading blackout with the minimum transmission power remaining in the system. The critical elements and their IDs can be determined by the vector $F(u(k))$ once the optimal control input vector $u(k)$ is obtained.

\end{remark}
\begin{remark}
From the DOP formulation presented above, we can see that the external disturbances are only applied in the first step ($k=0$), that is, the initial disturbance vector $u(0)$. The DOP formulation can be extended to the case where external disturbances or control inputs are applied in different steps, which helps facilitate future studies on human errors in cascading failures and protection reactions.
\end{remark}

\section{DOP Solution }\label{sec:ca}

The DOP can essentially be viewed as a control problem, where we search for an optimal control input vector $u(k)$ to pin the power gird to certain worst-case cascading blackout defined in Eq.\,(10). In this section, the ICRA based on the maximum principle of the optimal control theory is applied to solve the DOP as presented in Eqs.\,(9), (10) and (11). The maximum principle is a powerful method for the computation of optimal controls, with the crucial advantages that it does not require prior evaluation of the infimal cost function and provides necessary conditions for optimality of solutions. In the following, the Lagrange multiplier method in the maximum principle is presented in detail. 

Introduce the Lagrange multipliers $[{\lambda_{k+1}}]\triangleq [{\lambda_{1},...,\lambda_{N}}]$, $\lambda_{k+1}\in \mathbb{R}^{n}$ (usually referred to as adjoint variables) to the Eqs.\,(9), (10) and (11). The Lagrangian function shall be as follows:
\begin{equation}
\begin{aligned}
\mathfrak{L}(Y_P, \lambda)\triangleq ||P^{N}(Y_P^{N})||_{F}^{2}+\epsilon\sum\limits_{k=0}^{N_c-1} [\frac{1}{\max\left\{0,1-k\right\}}\times\frac{1}{\max^2\left\{0,N_n-||F(u_k)||^{2}\right\}}]\\
+\lambda_{k+1}^{T}\left\{G[P_{ij}^{k}(Y_P^{k}),\ C_{tr1}]Y_P^{k}+Diag[-u(k)]F(u_{k})-Y_P^{k+1}\right\}
\end{aligned}
\end{equation}
where $\lambda\triangleq[\lambda_{1}^{T}\quad \lambda_{1}^{T}...\quad \lambda_{N}^{T}]^T$.

To guarantee the existence of the partial derivative $\partial Y_{P}^{k+1}/\partial Y_{P}^{k}$, hereafter we make the assumption that each sub-network that is isolated due to redistributions of power flows in the cascading process, the partial derivative $\partial Y_{P}^{k+1}/\partial Y_{P}^{k}$ can be non-singular or reduced-order non-singular on $\mathbb{R}^{n}\times \Omega$ \cite{26}. 

Let 
\begin{equation*}
Y_{P}^{*}\triangleq[(Y_{0}^{*})^T\quad... \quad(Y_{N}^{*})^T\quad (u_{0}^{*})^T\quad... \quad(u_{N-1}^{*})^T]^T
\end{equation*}
be the minimising vector corresponding to the sequences $[(Y_{0}^{*})\quad... \quad(Y_{N}^{*})]$ and $[u_{0}^{*}\quad... \quad u_{N-1}^{*}]$. Observe that the dual feasibility condition in the Karush-Kuhn-Tucker (KKT) optimality conditions is equivalent to the statement that there exists $\lambda^{*}\triangleq[(\lambda_{1}^{*})^{T}\quad (\lambda_{2}^{*})^{T}...\quad (\lambda_{N}^{*})^{T}]^T$ such that the partial derivative $\partial \mathfrak{L}/\partial Y_{P}$ of the Lagrangian function vanishes at $(Y_{P}^{*},\lambda^{*})$. 

Therefore, there hold the following conditions
\begin{equation}
\left\{
   \begin{array}{l l}
       \frac{\partial\mathfrak{L}(Y_{P}^{*},\lambda^{*})}{\partial Y_P^{k}}=0 \\
      \\ \frac{\partial\mathfrak{L}(Y_{P}^{*},\lambda^{*})}{\partial u_{k}}=0
       \end{array}
       \right. 
\end{equation}
where $\frac{\partial\mathfrak{L}}{\partial Y_P^{k}}$ and $\frac{\partial\mathfrak{L}}{\partial u_{k}}$ denote the row vectors of partial derivatives 
\begin{equation*}
\left\{
   \begin{array}{l l}
      \frac{\partial\mathfrak{L}}{\partial Y_P^{k}}\triangleq [\frac{\partial\mathfrak{L}}{\partial Y_{P1}^{k}}\quad ...\quad \frac{\partial\mathfrak{L}}{\partial Y_{PN}^{k}}]\\
 \\\frac{\partial\mathfrak{L}}{\partial u_{k}}\triangleq [\frac{\partial\mathfrak{L}}{\partial u_{k}^{1}}\quad ...\quad \frac{\partial\mathfrak{L}}{\partial u_{k}^{m}}]
       \end{array}
       \right. 
\end{equation*}
To perform the differentiations above, the Hamiltonian $\mathbb{H}$: $\mathbb{R}^n\times\mathbb{R}^m\times\mathbb{R}^n\rightarrow\mathbb{R}$  defined as follows is introduced
\begin{equation}
\begin{aligned}
\mathbb{H}(Y_P^{k},u_{k}, \lambda_{k})\triangleq \epsilon\sum\limits_{k=0}^{N_c-1} [\frac{1}{\max\left\{0,1-k\right\}}\times\frac{1}{\max^2\left\{0,N_n-||F(u_k)||^{2}\right\}}]\\
+\lambda_{k+1}^{T}\left\{G[P_{ij}^{k}(Y_P^{k}),\ C_{tr1}]Y_P^{k}+Diag[-u(k)]F(u_{k})\right\}
\end{aligned}
\end{equation}
where the first term of the Hamiltonian $\mathbb{H}$ (denoted as $L$) is the per-stage weighting in the cost function.

Note that
\begin{equation*}
\left\{
   \begin{array}{l l}
  \frac{\partial\mathbb{H}}{\partial Y_P^{k}}= \frac{\partial L}{\partial Y_{P}^{k}}+\lambda_{k}^{T}\frac{\partial Y_{P}^{k+1}}{\partial Y_{P}^{k}}\\
\\ \frac{\partial\mathbb{H}}{\partial u_{k}}= \frac{\partial L}{\partial u_{k}}+\lambda_{k}^{T}\frac{\partial Y_{P}^{k+1}}{\partial u_{k}}
       \end{array}
       \right. 
\end{equation*}
where 
\begin{equation*}
\left\{
   \begin{array}{l l}
      \frac{\partial L}{\partial Y_{P}^{k}}\triangleq [\frac{\partial L}{\partial Y_{P1}^{k}}\quad ...\quad \frac{\partial L}{\partial Y_{PN}^{k}}]\\
 \\\frac{\partial L}{\partial u_{k}}\triangleq [\frac{\partial L}{\partial u_{k}^{1}}\quad ...\quad \frac{\partial L}{\partial u_{k}^{m}}]
       \end{array}
       \right. 
\end{equation*}
Thus, the following conditions hold 
\begin{equation}
\frac{\partial\mathfrak{L}(Y_P^{*}, \lambda^{*})}{\partial Y_P^{k}}=\frac{\partial\mathbb{H}(Y_P^{k*},(u_{k}^{*}),\lambda_{k+1}^{*})}{\partial Y_P^{k}}-(\lambda_{k}^{*})^T=0
\end{equation}
\begin{equation}
\frac{\partial\mathfrak{L}(Y_P^{*}, \lambda^{*})}{\partial Y_P^{N}}=\frac{\partial(||P^{N}(Y_P^{N})||_{F}^{2})}{\partial Y_P^{k}}-(\lambda_{N}^{*})^T=0
\end{equation}
\begin{equation}
\frac{\partial\mathfrak{L}(Y_P^{*}, \lambda^{*})}{\partial u_{k}}=\frac{\partial\mathbb{H}(Y_P^{k*},(u_{k}^{*}),\lambda_{k}^{*})}{\partial u_{k}}=0
\end{equation}
Further, the following equations can be obtained
\begin{itemize}
\item[(i)] State equation:
\begin{equation}
Y_P^{(k+1)*}=G[P_{ij}^{k}(Y_P^{k*}),\ C_{tr1}]Y_P^{k}+Diag[-u(k)]F^{*}(u_{k}) 
\end{equation}
\end{itemize}

\begin{itemize}
\item[(ii)] Adjoint equation:
\begin{equation}
(\lambda_{k}^{*})^T=\frac{\partial\mathbb{H}(Y_P^{k*},F^{*}(u_{k}),\lambda_{k+1}^{*})}{\partial Y_P^{k}}
\end{equation}
\end{itemize}

\begin{itemize}
\item[(iii)] Boundary equation:
\begin{equation}
(\lambda_{N}^{*})^T=\frac{\partial(||P^{N}(Y_P^{N})||_{F}^{2})}{\partial Y_P^{N}}
\end{equation}
\end{itemize}

\begin{itemize}
\item[(iv)] Hamiltonian condition:
\begin{equation}
\frac{\partial\mathbb{H}(Y_P^{k*},u_{k}^{*},\lambda_{k+1}^{*})}{\partial u_{k}}=0
\end{equation}
\end{itemize}

We show the main steps for solving Eqs.\,(18), (19), (20) and (21).

First, we solve the adjoint equations. From Eqs.\,(14) and (19), the following equation can be obtained
\begin{equation}
\lambda_{k}^{*}=(\frac{\partial Y_P^{k+1}}{\partial Y_P^{k}})^T\lambda_{k+1}^{*}
\end{equation}
where the dimension of $\frac{\partial Y_P^{k+1}}{\partial Y_P^{k}}$ is $N\times N$. From Eq.\,(11), we know that each element of $Y_P^{k+1}$ can be determined by
\begin{equation}
y_{P,i}^{k+1}=g_{i}(p_{ij}^{k},\ C_{tr1i})y_{P,i}^{k}+Diag[-u(k)]_{ii}f_{i}(u_{k})
\end{equation}
Hence, from Eqs.\,(22) and (23), the following partial derivative can be obtained
\begin{equation}
\frac{\partial y_{P,i}^{k+1}}{\partial y_{P,i}^{k}}=\frac{\partial g_{i}}{\partial p_{ij}^{k}}\cdot \frac{\partial p_{ij}^{k}}{\partial y_{P,is}^{k}}\cdot y_{P,i}^{k}+g_{i}(p_{ij}^{k},\ C_{tr1i})\quad s=1,2,...,i,...,N_c
\end{equation}
where the partial derivative of $\frac{\partial p_{ij}^{k}}{\partial y_{P,is}^{k}}$ is the constant zero except when $s=i$. The term $\frac{\partial g_{i}}{\partial P_{ij}^{k}}$ is as follows:
\begin{eqnarray}\frac{\partial g_{i}}{\partial P_{ij}^{k}}=
\begin{cases}
-a\cdot p_{ij}^{k}{\rm{cos}}a(|p_{ij}^{2}|-C_{tr1i}^{2}), &\sqrt{C_{tr1i}^{2}-\frac{\pi}{2a}}\leq |p_{ij}| \leq \sqrt{C_{tr1i}^{2}+\frac{\pi}{2a}} \cr 0, &\rm{otherwise} \end{cases}
\end{eqnarray}
and the term $\frac{\partial p_{ij}^{k}}{\partial y_{P,is}^{k}}$ is unknown which will be determined in the later part.

Second, we determine the boundary equation. The DC power flow equations are incorporated into the cascading failure model in Eq.\,(7) to get the expression of active power function as follows:
\begin{equation}
P_{ij}^{k}=(Ae_{i})^{T}Diag(Y_{P}^{k})(Ae_{j})(e_{i}-e_{j})^{T}[A^{T}Diag(Y_{P}^{k})A]^{-1}P
\end{equation}
where the vector of the active power $P$ is known for each iteration step. Meanwhile, the expression of the active power in the final step can be determined by
\begin{equation*}
P_{ij}^{N}=(Ae_{i})^{T}Diag(Y_{P}^{N})(Ae_{j})(e_{i}-e_{j})^{T}[A^{T}Diag(Y_{P}^{N})A]^{-1}P
\end{equation*}
 From Eqs.\,(20) and (26), the following equations can be obtained
 \begin{equation}
(\lambda_{N}^{*})^T=\frac{\partial[{{\sum\limits_{i=1}^N\sum\limits_{j=1}^N}(P^{N}_{ij})]^2}}{\partial Y_P^{N}}={{\sum\limits_{i=1}^N\sum\limits_{j=1}^N}2P^{N}_{ij}\frac{\partial p_{ij}^{N}}{\partial y_{P}^{N}}}
\end{equation}
\begin{equation}
\begin{aligned}
\frac{\partial p_{ij}^{N}}{\partial y_{P,is}^{N}}=(Ae_{i})^{T}\frac{\partial Diag(Y_{P}^{N})}{\partial y_{P,is}^{N}}(Ae_{j})(e_{i}-e_{j})^{T}[A^{T}Diag(Y_{P}^{N})A]^{-1}P\\
+(Ae_{i})^{T}Diag(Y_{P}^{N})(Ae_{j})(e_{i}-e_{j})^{T}\frac{\partial [A^{T}Diag(Y_{P}^{N})A]^{-1}}{\partial y_{P,is}^{N}}P
\end{aligned}
\end{equation}
For simplicity, the matrix $\mathbf{E_{ii}}$ is used to represent the term $\frac{\partial Diag(Y_{P}^{N})}{\partial y_{P,is}^{N}}$. Then Eq.\,(28) can be transferred into 
\begin{equation}
\begin{aligned}
\frac{\partial p_{ij}^{N}}{\partial y_{P,is}^{N}}&=(Ae_{i})^{T}\mathbf{E_{ii}}(Ae_{j})(e_{i}-e_{j})^{T}[A^{T}Diag(Y_{P}^{N})A]^{-1}P\\
&+(Ae_{i})^{T}Diag(Y_{P}^{N})(Ae_{j})(e_{i}-e_{j})^{T}\cdot [[{-A^{T}Diag(Y_{P}^{N})A]^{-1}\cdot A^{T} \mathbf{E_{ii}}A\cdot [A^{T}Diag(Y_{P}^{N})A]^{-1}}]P
\end{aligned}
\end{equation}
Finally, we determine the Hamiltonian condition 

\begin{equation}
\epsilon\frac{\partial [\frac{1}{\max\left\{0,1-k\right\}}\times\frac{1}{\max^{2}\left\{0,N_n-||F^{*}(u_k)||^{2})\right\}}]}{\partial u_{k}}+\frac{\partial [\lambda_{k+1}^{T}Diag[-u(k)]F^{*}(u_{k})]}{\partial u_{k}}=0
\end{equation}
where $F^{*}(u_{k})$ and $u_{k}$ are vectors with an $N\times1$ dimension. The following equation can be acquired according to Eq.\,(30)
\begin{equation}
\frac{4\epsilon}{\max\left\{0,1-k\right\}\times\max^{3}\left\{0,N_n-||F^{*}(u_k)||^{2}\right\}}[F^{*}(u_{k})]^{T}\frac{\partial F^{*}(u_k)}{\partial u_{k}}-\lambda_{k+1}^{T}\left\{Diag[-F(u(k)]+\frac{\partial F^{*}(u_k)}{\partial u_{k}}\right\}=0
\end{equation}
where the term $\frac{\partial F^{*}(u_k)}{\partial u_{k}}$ is as follows:
\[
  \frac{\partial F^{*}(u_k)}{\partial u_{k}}=
\begin{bmatrix}
\frac{\partial f^{*}(u_{k1})}{\partial u_{k1}}& 0 &\ldots& 0  \\
 0&\frac{\partial f^{*}(u_{k2})}{\partial u_{k2}}&\ldots&0  \\
\vdots&\vdots
 & \ddots & \vdots \\
 0& 0& \ldots & \frac{\partial f^{*}(u_{kn})}{\partial u_{kn}}
\end{bmatrix}
\]
and the term $\frac{\partial f_{i}^{*}(u_{k})}{\partial u_{ki}}$ is 
\begin{eqnarray*}\frac{\partial f_{i}^{*}(u_{k})}{\partial u_{ki}}=
\begin{cases}
bu_{ki}{\rm{cos}}b(|u_{ki}^{2}|-C_{tr2i}^{2}), &\sqrt{C_{tr2i}^{2}-\frac{\pi}{2b}}\leq |u_{ki}| \leq \sqrt{C_{tr2i}^{2}+\frac{\pi}{2b}} \cr 0, &\rm{otherwise} \end{cases}
\end{eqnarray*}

Now, necessary optimality conditions for the control input to be the minimisers of the optimization problem is derived. From Eqs.\,(20) and (22), the recursion formula for Lagrange multipliers is determined as
\begin{equation}
\lambda_{k+1}=\prod_{s=0}^{N-k-2}(\frac{\partial Y_{P}^{N-s}}{\partial Y_{P}^{N-s-1}})^{T}\lambda_{N}\quad s=1,2,...,N_c-1
\end{equation}
The solution of DOP can be determined by the following equations 
\begin{equation}
\left\{
   \begin{array}{l l}
  \frac{4\epsilon}{\max\left\{0,1-k\right\}\times\max^{3}\left\{0,N_n-||F^{*}(u_k)||^{2}\right\}}[F(u_{k})]^{T}\frac{\partial F(u_k)}{\partial u_{k}}-\prod\limits_{s=0}^{N-k-2}(\frac{\partial Y_{P}^{N-s}}{\partial Y_{P}^{N-s-1}})\lambda_{N}^{T}\left\{Diag[-F(u(k)]+\frac{\partial F(u_k)}{\partial u_{k}}\right\}=0\\
 \\Y_P^{k+1}=G[P_{ij}^{k}(Y_P^{k}),\ C_{tr1}]Y_P^{k}+Diag[-u(k)]F(u_{k})
       \end{array}
       \right. 
\end{equation}
where $Y_{P}^{k}$ and $u(k)$ are two unknown variables. The algorithm on identifying critical risk of cascading failures in power systems, denoted as ICRA, is summarized in the Table\,1. 

\begin{table}
 \caption{\label{ISA} Identifying Critical Risk Algorithm}
 \begin{center}
 \begin{tabular}{lcl} \hline
  1: Set the maximum step $i=i_{\max}$, $i=0$ and $J^*=J_{\min}$\\
  2: \textbf{while} ($i<=i_{\max}$) \\
  3: ~~~~~~~Solve the nonlinear algebraic equations in Eq.\,(33)  \\
  4: ~~~~~~~Determine the control input (external disturbances) $u(k)$\\
  5: ~~~~~~~Validate the control input $u(k)$ in Eq.\,(7)\\
  6: ~~~~~~~Compute the resulting cost function $J^i$ in Eq.\,(10) \\
  7: ~~~~~~~Determine the IDs of critical elements $F(u_k)$ in Eq.\,(8) \\
  8: ~~~~~~~\textbf{if} ($J^i>J^*$)  \\
 9: ~~~~~~~~~~~Set $u^*=u^i$ and $J^*=J^i$ \\
  10: ~~~~~~~\textbf{end if}  \\
 11: ~~~~~Set $i=i+1$ \\
 12: \textbf{end while} \\ \hline
 \end{tabular}
 \end{center}
\end{table}

\section{Simulation Results and Verification}

We consider two different cases for identifying the critical risks of cascading failures in power systems. 

\textbf{Case 1:} Both the critical elements and corresponding initial disturbances are unknown variables.

\textbf{Case 2:} The initial disturbances are given as branch outage where the critical elements remain to be identified. In this case, since the initial disturbances have to be element outages, we replace the vector $u(k)$ in Eq.\,(7) with the initial nodal admittance matrix $Y_P^{0}$. 

Note that the \textbf{Case 2} above is a special case of \textbf{Case 1}. We are particularly interested in this special case for two reasons: (i) in practice, branch outage is a common type of failures \cite{27}; and (ii) for this special case, the optimality of the solutions in small or medium-sized systems could be verified by brute force i.e., by considering all the possible combinations of  branch outage cases with a given number of outage branches. In simulations, we use $\emph{Matlab}$ with $\emph{fsolve}$ as the non-linear solver for solving $Y_{P}^{k}$ and $u_{k}$. For the test case data and calculation of the electric circuit parameters, the codes from $\emph{Matpower}$ are used extensively.

\subsubsection*{A. Simulation Results}

The test case of the IEEE 9-Bus system contains 3 generators, 6 branches, 3 loads and 2 winding power transformers. The test case of the IEEE 14-Bus system consists of 14 buses, 5 generators and 11 loads. Information about these two test cases and algorithmic parameters are presented in Table\,2.

\begin{table*}[!htbp]
\centering
\caption{Information of Two Test cases}\label{tab:aStrangeTable}
\begin{tabular}{ccc}
\toprule
Information& Test Case 1 (9-Bus)&Test Case 2 (14-Bus)\\

\midrule
Filename (in \emph{Matpower)}& case9.m&case14.m\\
Number of Nodes& 9&14\\
Number of Elements $N_e$& 9&20\\
Iteration Steps $N_c$&10&14\\
Weight of Cost Function $\epsilon$&0.02&0.02\\
Parameter $a$ and $b$& $5\times 10^{{5}}$&$8\times 10^{{5}}$\\
Threshold Value Vector $Ctr2$  & $Y_{p1}$&    $Y_{p2}-4E_{14\times1}$\\
Initial Value Vector of Solver &$-10*rand(N_e,N_c+1)$&$-9*rand(N_e,N_c+1)$\\

\bottomrule
\end{tabular}
\end{table*}
Therein, $Y_{p1}=[-17.36, -10.87, -5.88, -17.06, -9.92, -13.89, -16.00, -6.21]$ and $Y_{p2}=[-16.90, \\-4.48, -5.05, -5.67, -5.75, -5.85, -23.75 -4.78, -1.80, -3.97, -5.03, -3.91, -7.68, -5.68, -9.09, \\-11.83,  -3.70, -5.21, -5.00, -2.87]$ are the initial susceptance vectors of elements in per-unit for the 9-Bus and the 14-Bus test systems respectively. The threshold value vector $Ctr1$ for the 9-Bus and the 14-Bus test systems are $[0.8,1.8, 1.0, 0.5, 0.5, 1.0, 0.8, 0.7 ,0.5]$ and $[1.8,1.0,1.0,0.8,0.6,\\0.5,0.9,0.5,0.4,0.7,0.1,0.1,0.3,0.1,0.6,0.1,0.2,0.2,0.1]
$, respectively.

We first carry out simulations of \textbf{Case 1}. For the IEEE 9-Bus test system, the number of critical elements $N_n$ is set as 1. For the IEEE 14-Bus test system, the number of critical elements $N_n$ is set as 1 and 2. The results are presented in Table\,3.

\begin{table*}[!htbp]
\centering
\caption{Results of identifying the critical elements and corresponding initial disturbances}\label{tab:aStrangeTable}
\begin{tabular}{ccc}
\toprule
& IDs of the critical elements&Initial disturbances (p.u.)\\
\midrule
Test Case 9-Bus ($N_n=1$)& 1&17.36\\
Test Case 14-Bus ($N_n=1$) & 3&4.73\\
Test Case 14-Bus ($N_n=2$)&2 and 3&4.23 and 4.74\\
\bottomrule
\end{tabular}
\end{table*}

We then conduct simulations for \textbf{Case 2}. For the IEEE 9-Bus test system, the number of critical elements $N_n$ is set as 1. The identified critical element marked with the red oval is shown in the Fig. 1. For the IEEE 14-Bus test system, the number of critical elements $N_n$ is set as 1, 2, 3 or 4, respectively. The results are shown in Fig. 2. 

\begin{figure}
\scalebox{0.3}[0.3]{\includegraphics{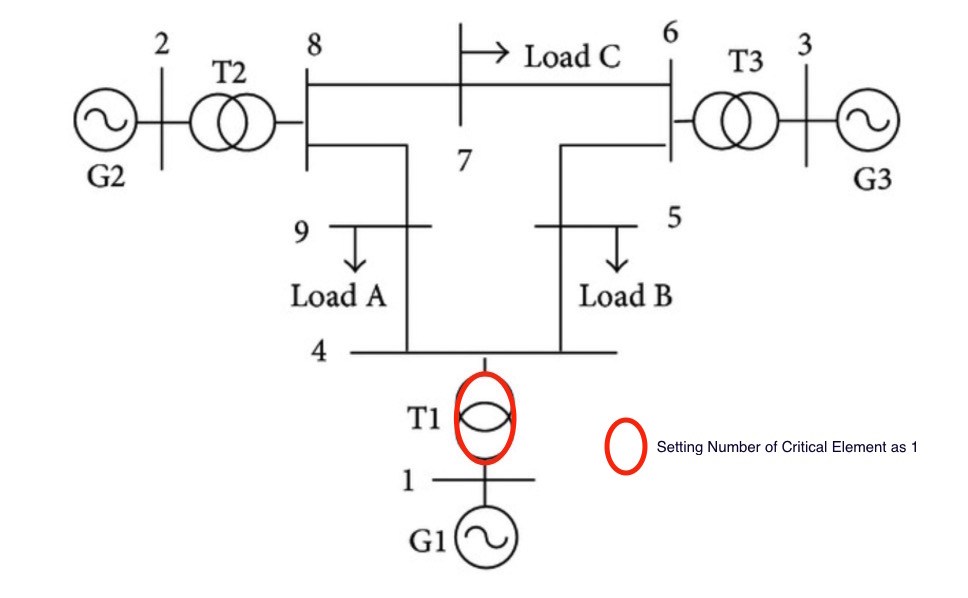}}\centering
\caption{\label{mg} The identified critical element (marked with red oval) in the test case of the IEEE-9 Bus system. The number of critical elements is set as 1.}
\end{figure}

\begin{figure}
\scalebox{0.3}[0.3]{\includegraphics{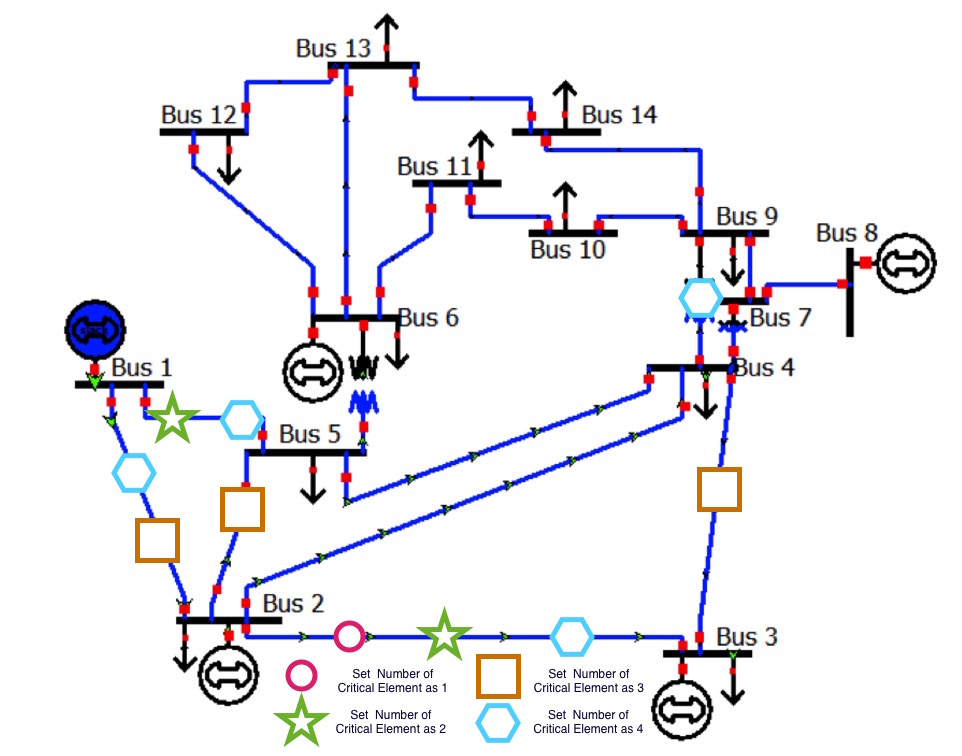}}\centering
\caption{\label{mg} The identified critical elements for the test case of the IEEE-14 Bus system when the number of critical elements is set as 1, 2, 3 or 4, respectively.}
\end{figure}

%\begin{remark}

%In simulations, we obtain some other results in \textbf{Case 1} and \textbf{Case 2} for the two test systems. For example, for \textbf{Case 1} in the IEEE-9 Bus test system, applying the initial disturbance $u=3.12$ to the $8^{th}$ element (connected between bus 6 and bus 9) leads to outage of eight branches; for \textbf{Case 2} in the IEEE-9 Bus test system, the possible ID of the critical element is 1 (connected between bus 1 and bus 4), or 2 (connected between bus 2 and bus 7) or 3 (connected between bus 3 and bus 9) element; for \textbf{Case 1} in the IEEE-14 Bus test system, applying the initial disturbances $u_3=4.92$ and $u_{19}=2.60$ to the $3^{rd}$ element and the $19^{th}$ element (connected between bus 12 and bus 13) respectively, the transmission power is 0.17 p.u. and so forth. All these results give rise to widespread disruption to power systems, which are beneficial information for power system planners.
%\end{remark}

\subsubsection*{B. Verification}

In this subsection, the correctness of the numerical results generated by the ICRA, as reported in Subsection A, are verified. For verifying \textbf{Case 1}, the computed initial disturbances are applied to the corresponding elements (see Table\,3) in the two test systems respectively. For verifying \textbf{Case 2}, the optimality of the solution can be verified by brute force, i.e., by considering all the possible combinations of  branch outage cases with a given number of outage branches. More specifically, the cascading failure model in Eq.\,(7) and the DC power-flow model in Eq.\,(11) from the ICRA are used extensively, and the numerical simulation results on the critical elements and disruptive disturbances are validated by disturbing the selected element with the computed magnitude of disturbances in the corresponding IEEE Bus test systems. The final remaining transmission power and/or the final network topology is proposed to quantify the disruption. In the following, the verification results are given.

For \textbf{Case 1} in the IEEE 9-Bus test system, we apply the corresponding initial disturbance, that is, $u=17.36$ to the element 1 connected between bus 1 and bus 4. The corresponding initial disturbance $u=17.36$ equals the outage of the element 1. The evolution process of transmission network topology is shown in Fig.\,3. For the simulation results of \textbf{Case 2}, the initial disturbances take form of element outage. After testing all the possible element outage cases, we find all branches (elements) finally are broken when the element 1 is taken down, which has the same result with \textbf{Case 1} . 

\begin{figure}
\scalebox{0.4}[0.4]{\includegraphics{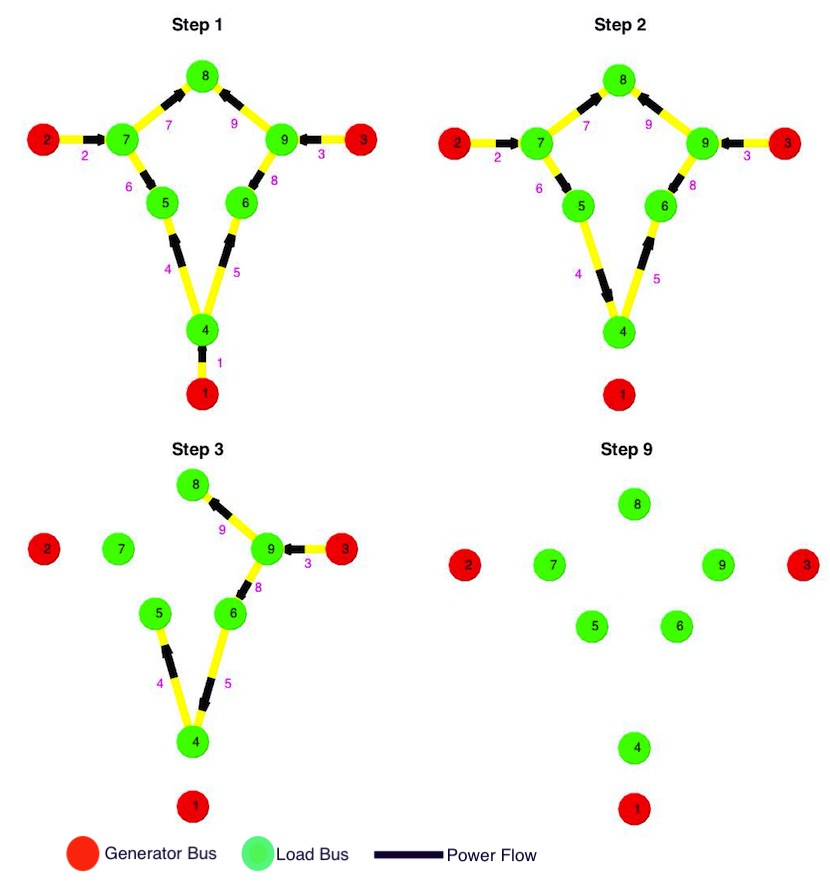}}\centering
\caption{\label{mg} The diagram of propagation process of cascading failure and final power grid topology with the element 1 being broken as the initial disturbance in the IEEE 9-Bus test system. }
\end{figure}

As we can see from Fig.\,3, with the element 1 being broken as the initial disturbance, all the branches (elements) are broken and the final remaining transmission power becomes zero. The verification results are the same to the results presented in Table 3 and Fig.\,1, which verifies the correctness of proposed algorithm of ICRA. 

For the IEEE 14-Bus test system, the initial power transmission is 3.07 p.u. when the power system operates in normal status. For the simulation results of \textbf{Case 1}, when the number of critical elements $N_n$ is set as 1, we apply the corresponding initial disturbance, that is, $u=4.73$ to the element 3 connected between bus 2 and bus 3. The remaining transmission power is 0.02 p.u. When $N_n$ is set as 2, we apply the initial disturbances $u_2=4.23$ and $u_3=4.73$ to the element 2 (connected between bus 1 and bus 5) and the element 3 respectively. The remaining transmission power becomes zero. For the simulation results of \textbf{Case 2}, we apply outage of one element, two elements, three elements and four elements respectively as the initial contingencies. We simulate all possible element outage cases when the number of outage elements varies from one to four in the IEEE 14-Bus test system; the results are shown in Fig.\,4.

\begin{figure}\centering
 {\includegraphics[width=0.42\textwidth]{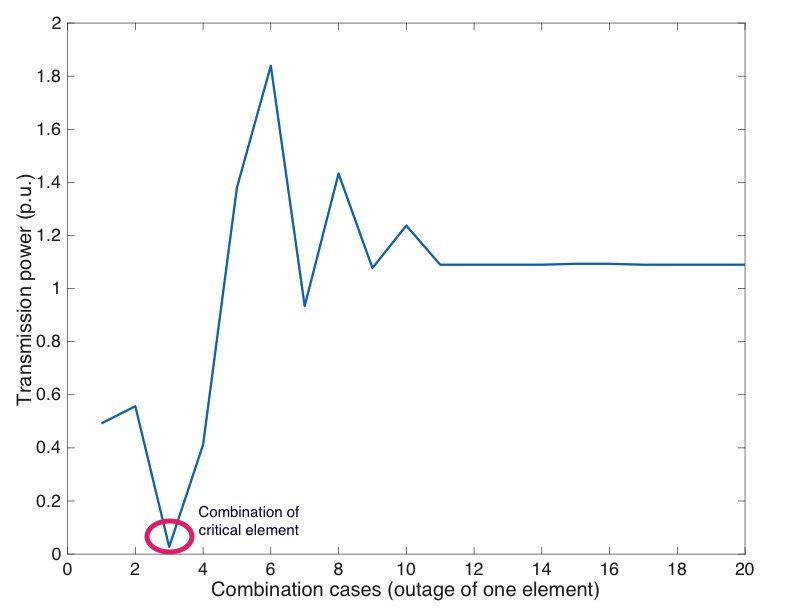}}
 {\includegraphics[width=0.42\textwidth]{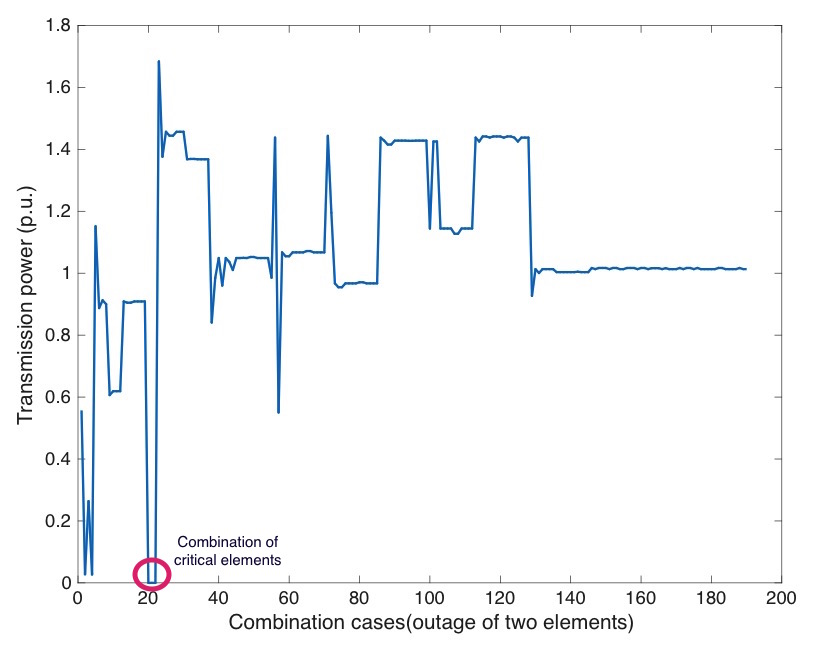}} \\
 {\includegraphics[width=0.42\textwidth]{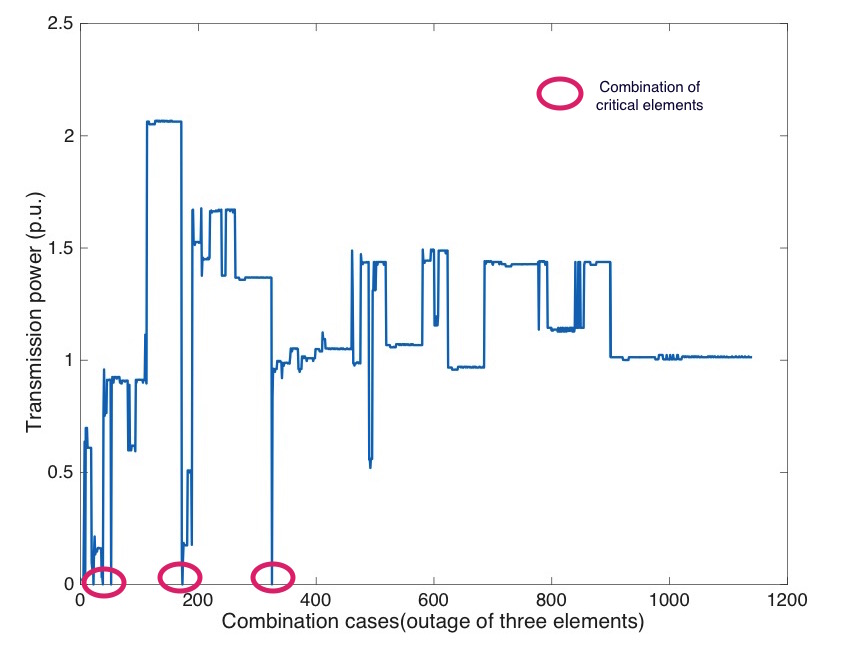}}
 {\includegraphics[width=0.42\textwidth]{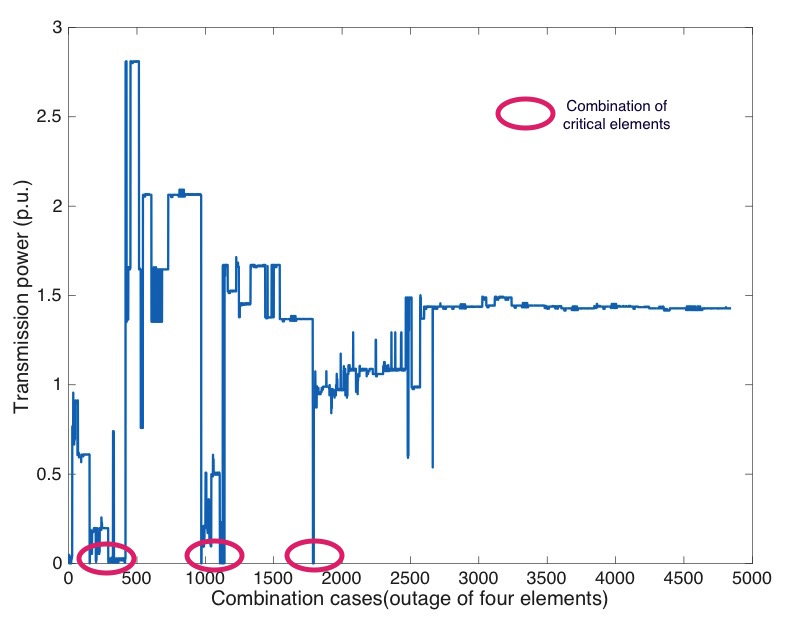}}
 \caption{\label{id9} The transmission power (in p.u.) left when the number of outage elements varies from one to four in the IEEE 14-Bus test system and the combination of critical elements are circled with read ovals.}
\end{figure}

As we can see from the Fig.\,4, the outage of the element 3 results in the minimum remaining transmission power, that is 0.02 p.u. When $N_n=2$,  the combinations (IDs of elements) $[2,3]$, $[2,4]$ and $[2,5]$ lead to zero transmission power. The combination $[1,5,6]$ results in zero transmission power when $N_n=3$. When $N_n=4$, the combinations [1,2,3,9], [1,2,3,10], [1,2,3,11], [1,2,3,12], [1,2,3,13], [2,4,6,7], [2,4,6,8], [2,4,6,9], [2,4,6,10], [2,4,6,11], [2,4,6,12] and [2,4,6,13] lead to zero transmission power. From the verification results of \textbf{Case 1} and \textbf{Case 2} above for the IEEE 14-Bus test system, we can verify the correctness of the simulation results in Table 3 and Fig.\,2.	

From the simulation and verification results on the IEEE 9-Bus and the IEEE 14-Bus test systems, we may conclude that the proposed ICRA is effective. 

\section{Conclusions and Future Work}\label{sec:conclusions}

In this paper, the problem of identifying critical risks of cascading failures in power systems was formulated as a dynamic optimization problem (DOP) within the framework of optimal control theory. By pinning the power system into the worst-case cascading blackout, the optimal control inputs that reflect the critical elements and corresponding disturbances were determined by solving the DOP. The ICRA based on the maximum principle was applied to solve the DOP, which provides the necessary conditions for optimality of solutions. The correctness and effectiveness of the ICRA have been verified by applying the computed initial disturbances or elements outage to the corresponding elements in IEEE Bus systems. The efficient identification of critical risks may help power system planners to reveal hidden catastrophic risks, preplan system protection and recovery, and consequently improve system resilience. The research work will be extended to include identifying critical risks as disturbances to network nodes and other mechanisms such as generator tripping, load shedding and voltage collapse, etc. In a longer term, we shall take into account the cost of protection and recovery while identifying the worst cases.

\section*{Acknowledgement}

This study is an outcome of the Future Resilient System (FRS) project at the Singapore-ETH Centre (SEC), which is funded by the National Research Foundation of Singapore (NRF) under its Campus for Research Excellence and Technological Enterprise (CREATE) program. Part of this work is also supported by the Ministry of Education (MOE), Singapore, under Contract No. MOE 2016-T2-1-119.

\end{document}